\documentclass[12pt,twoside]{article}
\usepackage{amssymb}
\usepackage{amsmath}
\usepackage{mathrsfs} 
\usepackage{amsfonts}
\usepackage[dvips]{graphicx}
\begin{document}
\newtheorem{d1}{Definition}[section]
\newtheorem{c1}{Corollary}[section]
\newtheorem{l1}{Lemma}[section]
\newtheorem{r1}{Remark}[section]

\newcommand{\cA}{{\cal A}}
\newcommand{\cB}{{\cal B}}
\newcommand{\cC}{{\cal C}}
\newcommand{\cD}{{\cal D}}
\newcommand{\cE}{{\cal E}}
\newcommand{\cF}{{\cal F}}
\newcommand{\cG}{{\cal G}}
\newcommand{\cH}{{\cal H}}
\newcommand{\cI}{{\cal I}}
\newcommand{\cJ}{{\cal J}}
\newcommand{\cK}{{\cal K}}
\newcommand{\cL}{{\cal L}}
\newcommand{\cM}{{\cal M}}
\newcommand{\cN}{{\cal N}}
\newcommand{\cO}{{\cal O}}
\newcommand{\cP}{{\cal P}}
\newcommand{\cQ}{{\cal Q}}
\newcommand{\cR}{{\cal R}}
\newcommand{\cS}{{\cal S}}
\newcommand{\cT}{{\cal T}}
\newcommand{\cU}{{\cal U}}
\newcommand{\cV}{{\cal V}}
\newcommand{\cX}{{\cal X}}
\newcommand{\cW}{{\cal W}}
\newcommand{\cY}{{\cal Y}}
\newcommand{\cZ}{{\cal Z}}

\def\benrr{\begin{eqnarray*}}
\def\eenrr{\end{eqnarray*}}\def\bc{\begin{center}}
\def\ec{\end{center}}
\def\d{\dot}
\def\si{\sigma}

\def\fr{\frac}
\def\sq{\sqrt}

\def\et{\end{tabular}}
\def\la{\lambda}
\def\lan{\langle}
\def\ran{\rangle}

\bc

\textbf{ SEPARABILITY CRITERION FOR MULTIPARTITE QUANTUM STATES BASED ON THE BLOCH REPRESENTATION OF DENSITY MATRICES}

\vspace{.2 in} 

 \textbf{ ALI SAIF M. HASSAN\footnote{ Electronic address: alisaif@physics.unipune.ernet.in} and  PRAMOD S. JOAG\footnote{ Electronic address: pramod@physics.unipune.ernet.in}}\\
 Department of Physics, University of Pune, Pune, India-411007.
 
\ec

\vspace{.2in}

We give a new separability criterion, a necessary condition for separability of $N$-partite quantum states. The criterion is based on the Bloch representation of a $N$-partite quantum state and makes use of multilinear algebra, in particular, the matrization of tensors. Our criterion applies to {\it arbitrary} $N$-partite quantum states in $\mathcal{H}=\mathcal{H}^{d_1}\otimes \mathcal{H}^{d_2} \otimes \cdots \otimes \mathcal{H}^{d_N}.$ The criterion can test whether a $N$-partite state is entangled and can be applied to different partitions of the $N$-partite system.  We provide examples that show the ability of this criterion to detect entanglement. We show that this criterion can detect bound entangled states. We prove a sufficiency condition for separability of a 3-partite state, straightforwardly generalizable  to the case  $N > 3,$ under certain  condition. We also give a necessary and sufficient condition for separability of a class of $N$-qubit states which includes $N$-qubit PPT states.

\vspace{.2in}
{\it Keywords} : Bloch representation of quantum state, separability criteria, matrization of tensors, PPT entangled states.\\

\vspace{.2in}
\section{ Introduction}

 The question of quantifying entanglement of multipartite quantum states is fundamental to the whole field of quantum information and in general to the physics of multicomponent quantum systems. 
 Whereas entanglement of pure bipartite states is well understood, the classification of mixed states according to the degree and character of their entanglement is still a matter of intense research [1,2]. A $N$-partite state acting on 
 $\mathcal{H}=\mathcal{H}^{d_1}\otimes \mathcal{H}^{d_2} \otimes \cdots \otimes \mathcal{H}^{d_N}$ is separable [3] ( or fully separable) if it can be written as a convex sum of tensor products of subsystem states 
 $$\rho =\sum_w p_w \rho_w^{(1)}\otimes \rho_w^{(2)}\dots \otimes \rho_w^{(N)}=\sum_w p_w \bigotimes_{j=1}^N \rho_w^{(j)} , \; p_w > 0; \; \sum_w p_w=1 .\eqno{(1)}$$
 
 A state is called $k$ separable if we can write 
  $$\rho =\sum_w p_w \rho_w^{(a_1)}\otimes \rho_w^{(a_2)}\dots \otimes \rho_w^{(a_k)} \eqno{(2)}$$
   where  $a_i ; \; i=1,2,\dots ,k $ are the disjoint subsets of $\{1,2,\dots,N \}$ and $\rho^{(a_i)}$  acts on the tensor product space made up by the factors of $\mathcal{H}$ labeled by the members of $a_i$.
    The understanding of multipartite entanglement has progressed by dealing with some special classes of states like the density operators supported on the symmetric subspace of $\mathcal{H}$ [4]. A lower bound on concurrence on the multipartite mixed states is obtained [5]. K. Chen and L. Wu have given a generalized partial transposition and realignment criterion to detect entanglement of a multipartite quantum state [6]. 
     
 There are two definitions commonly used for the entanglement of multipartite quantum states, the one from Ref. [7] (ABLS) and the one introduced in [8] (DCT). In DCT, all possible partitions of N parties are considered and it is tested for each partition if the state is fully separable there or not. A state is called N partite entangled if it is not separable for any partition. If a state is separable for a bipartite partition, it is called biseparable.
In ABLS, a state is called biseparable if it is a convex combination of biseparable states, possibly concerning different partitions. A $N$-partite entangled state is one which is not biseparable. 
 
 In this paper we derive a necessary  condition for the  separability of multipartite quantum states for arbitrary finite  dimensions of the subsystem Hilbert spaces and without any further restriction on them. The criterion is based on the Bloch representation of a multipartite quantum state, which has been used in previous works to characterize the separability of bipartite density matrix, in particular, our work is a generalization of de Vicente's work on bipartite systems [9]. We make use of the algebra of higher order tensors, in particular the matrization of a tensor [10,11,12,13,14,15,16,17].
 
 The paper is organized  as follows. In section II we present the Bloch representation of a $N$-partite quantum state. In section III we obtain the main results on separability of a $N$-partite quantum state. In section IV we give a sufficient condition for the separability of a 3-partite quantum state generalizable to the case $N > 3$.
 In section V we investigate our separability criterion for mixed states, in particular, bound entangled states. 
 Finally we summarize in section VI. \\
  
\section{ Bloch Representation of a $N$-Partite Quantum State}
 
Bloch representation [18,19,20,21,22] of a density operator acting on the Hilbert space of a $d$-level quantum system $\mathbb{C}^d$ is given by [9] $$ \rho = \fr{1}{d} (I_d + \sum_i s_i \lambda_i) \eqno{(3)}$$ 
 Eq.(3) is the expansion of $\rho$ in the Hilbert-Schmidt basis $\{I_d,\lambda_i; i=1,2,\dots,d^2-1\}$ where $\lambda_i$ are the traceless hermitian generators of $SU(d)$ satisfying $Tr(\lambda_i \lambda_j)=2\delta_{ij}$
  and are characterized by the structure constants of the corresponding Lie algebra, $f_{ijk}, g_{ijk}$ which are,
     respectively, completely antisymmetric and completely symmetric.
 $$\lambda_i \lambda_j =\fr{2}{d} \delta_{ij} I_d + i f_{ijk}\lambda_k +g_{ijk}\lambda_k    \eqno{(4)}$$
 
 $\textbf{s} = (s_1,s_2,\dots,s_{d^2-1})$ in Eq.(3) are the vectors in $\mathbb{R}^{d^2-1}$, constrained by the positive semidefiniteness of $\rho$,  called Bloch vectors [21]. The set of all Bloch vectors that constitute a density operator is known as the Bloch vector space $B(\mathbb{R}^{d^2-1})$. The problem of determining $B(\mathbb{R}^{d^2-1})$ where  $d\ge 3$ is still open [19,20]. However, for pure states $(\rho=\rho^2)$ the following relations hold.
 $$||\textbf{s}||_2 = \sqrt\fr{d(d-1)}{2};\;\;\; s_i s_j g_{ijk}=(d-2)s_k     \eqno{(5)}$$
  where $||.||_2$ is the Euclidean norm in $\mathbb{R}^{d^2-1}$.
  
  It is known [23,24] that $B(\mathbb{R}^{d^2-1})$ is a subset of the ball $D_R(\mathbb{R}^{d^2-1})$ of radius $R=\sq{\fr{d(d-1)}{2}}$ , which is the minimum ball containing it, and that the ball $D_r(\mathbb{R}^{d^2-1})$ of radius $r=\sq{\fr{d}{2(d-1)}}$ is included in $B(\mathbb{R}^{d^2-1})$. That is, 
  
  $$D_r(\mathbb{R}^{d^2-1})\subseteq B(\mathbb{R}^{d^2-1}) \subseteq D_R(\mathbb{R}^{d^2-1}) \eqno{(6)}$$
  
  In order to give the Bloch representation of a density operator acting on the Hilbert   
  space $\mathbb{C}^{d_1} \otimes \mathbb{C}^{d_2} \otimes \cdots \otimes \mathbb{C}^{d_N}$
  of a $N$-partite quantum system, we introduce following notation. We use $k$, \; $k_i \; (i=1,2,\cdots)$ to denote a subsystem chosen from $N$ subsystems, so that $k$,\; $k_i \; (i=1,2,\cdots)$ take values in the set  $\mathcal{N}=\{1,2,\cdots,N\}$. The variables $\alpha_k \;\mbox{or} \; \alpha_{k_i}$ for a given $k$ or $k_i$ span the set of generators of $SU(d_k)\; \mbox{or}\;  SU(d_{k_i})$ group (Eqs.(3) and (4)) for the $k$th or $k_i$th subsystem, namely the set $\{\la_{1_{k_i}},\la_{2_{k_i}},\cdots,\la_{d_{k_i}^2-1}\}$ for the $k_i$th subsystem. For two subsystems $k_1$ and $k_2$ we define
  
   $$\lambda^{(k_1)}_{\alpha_{k_1}}=(I_{d_1}\otimes I_{d_2}\otimes \dots \otimes \lambda_{\alpha_{k_1}}\otimes I_{d_{k_1+1}}\otimes \dots \otimes I_{d_N})   $$
   $$\lambda^{(k_2)}_{\alpha_{k_2}}=(I_{d_1}\otimes I_{d_2}\otimes \dots \otimes \lambda_{\alpha_{k_2}}\otimes I_{d_{k_2+1}}\otimes \dots \otimes I_{d_N})  $$
   $$\lambda^{(k_1)}_{\alpha_{k_1}} \lambda^{(k_2)}_{\alpha_{k_2}}=(I_{d_1}\otimes I_{d_2}\otimes \dots \otimes \lambda_{\alpha_{k_1}}\otimes I_{d_{k_1+1}}\otimes \dots \otimes \lambda_{\alpha_{k_2}}\otimes I_{d_{k_2+1}}\otimes I_{d_N})   \eqno{(7)}$$
   
  where  $\lambda_{\alpha_{k_1}}$ and $\lambda_{\alpha_{k_2}}$ occur at the $k_1$th and $k_2$th places (corresponding to $k_1$th and $k_2$th subsystems respectively) in the tensor product and are the $\alpha_{k_1}$th and  $\alpha_{k_2}$th generators of $SU(d_{k_1}),\; SU(d_{k_2}) ,\alpha_{k_1}=1,2,\dots,d_{k_1}^2-1\; \mbox{and} \; \alpha_{k_2}=1,2,\dots,d_{k_2}^2-1$ respectively. Then we can write

$$\rho=\fr{1}{\Pi_k^N d_k} \{\otimes_k^N I_{d_k}+ \sum_{k \in \mathcal{N}}\sum_{\alpha_{k}}s_{\alpha_{k}}\lambda^{(k)}_{\alpha_{k}} +\sum_{\{k_1,k_2\}}\sum_{\alpha_{k_1}\alpha_{k_2}}t_{\alpha_{k_1}\alpha_{k_2}}\lambda^{(k_1)}_{\alpha_{k_1}} \lambda^{(k_2)}_{\alpha_{k_2}}+\cdots +$$
$$\sum_{\{k_1,k_2,\cdots,k_M\}}\sum_{\alpha_{k_1}\alpha_{k_2}\cdots \alpha_{k_M}}t_{\alpha_{k_1}\alpha_{k_2}\cdots \alpha_{k_M}}\lambda^{(k_1)}_{\alpha_{k_1}} \lambda^{(k_2)}_{\alpha_{k_2}}\cdots \lambda^{(k_M)}_{\alpha_{k_M}}+ \cdots+\sum_{\alpha_{1}\alpha_{2}\cdots \alpha_{N}}t_{\alpha_{1}\alpha_{2}\cdots \alpha_{N}}\lambda^{(1)}_{\alpha_{1}} \lambda^{(2)}_{\alpha_{2}}\cdots \lambda^{(N)}_{\alpha_{N}}\}.\eqno{(8)}$$

 where $\textbf{s}^{(k)}$ is a Bloch vector corresponding to $k$th subsystem,  $\textbf{s}^{(k)} =[s_{\alpha_{k}}]_{\alpha_{k}=1}^{d_k^2-1} $ which is a tensor of order one defined by
 $$s_{\alpha_{k}}=\fr{d_k}{2} Tr[\rho \lambda^{(k)}_{\alpha_{k}}]= \fr{d_k}{2} Tr[\rho_k \lambda_{\alpha_{k}}],\eqno{(9a)}$$ where $\rho_k$ is the reduced density matrix for the $k$th subsystem. Here $\{k_1,k_2,\cdots,k_M\},\; 2 \le M \le N,$ is a subset of $\mathcal{N}$ and can be chosen in $\binom{N}{M}$  ways, contributing $\binom{N}{M}$ terms in the sum $\sum_{\{k_1,k_2,\cdots,k_M\}}$ in Eq.(8), each containing a tensor of order $M$. The total number of terms in the Bloch representation of $\rho$ is $2^N$. We denote the tensors occurring in the sum $\sum_{\{k_1,k_2,\cdots,k_M\}},\; (2 \le M \le N)$ by $\mathcal{T}^{\{k_1,k_2,\cdots,k_M\}}=[t_{\alpha_{k_1}\alpha_{k_2}\cdots \alpha_{k_M}}]$ which  are defined by 
 
 $$t_{\alpha_{k_1}\alpha_{k_2}\dots\alpha_{k_M}}=\fr{d_{k_1}d_{k_2}\dots d_{k_M}}{2^M} Tr[\rho \lambda^{(k_1)}_{\alpha_{k_1}} \lambda^{(k_2)}_{\alpha_{k_2}}\cdots \lambda^{(k_M)}_{\alpha_{k_M}}]$$
 
$$ =\fr{d_{k_1}d_{k_2}\dots d_{k_M}}{2^M} Tr[\rho_{k_1k_2\dots k_M} (\lambda_{\alpha_{k_1}}\otimes\lambda_{\alpha_{k_2}}\otimes\dots \otimes\lambda_{\alpha_{k_M}})]   \eqno{(9b)}$$

where $\rho_{k_1k_2\dots k_M}$ is the reduced density matrix for the subsystem $\{k_1 k_2\dots k_M\}$. We call  The tensor in last term in Eq. (8) $\mathcal{T}^{(N)}$.\\



\section{ Separability Conditions}

 Before we obtain the main results we need following definition. Throughout the paper, we use the bold letter for vector and normal letter for components of a vector, matrix and tensor elements.\\ 

A rank-1 tensor is a tensor that consists of the outer product of a number of vectors. For $M$th order tensor $\mathcal{T}^{(M)}$ and $M$ vectors \; $\mathbf{u}^{(1)},\mathbf{u}^{(2)},\dots,\mathbf{u}^{(M)}$ this means that 
$t_{i_1i_2\dots i_M}=u_{i_1}^{(1)}u_{i_2}^{(2)}\dots u_{i_M}^{(M)}$
for all values  of the indices. This  is concisely written  as 
$\mathcal{T}^{(M)}=\mathbf{u}^{(1)}\circ \mathbf{u}^{(2)} \circ \dots \circ \mathbf{u}^{(M)}$[17,11].

Also, given two tensors $\mathcal{T}^{(M)}$  and $\mathcal{S}^{(N)}$ of  order $M$ and $N$ respectively, with 
 dimensions $I_1 \times I_2 \times \dots \times I_M$ and $J_1\times J_2 \times \dots\times J_N$ respectively, their outer product is defined as [16,10]
 
 $$(\mathcal{T}^{(M)}\circ \mathcal{S}^{(N)})_{i_1 i_2 \dots i_M j_1 j_2 \dots j_N}= t_{i_1 i_2\dots i_M}s_{j_1 j_2 \dots j_N} \eqno{(10)}$$
 
 \textbf{ Proposition 1 }: A  pure $N$-partite quantum state with Bloch representation (8) is fully separable (product state) if and only if      $$\mathcal{T}^{\{k_1,k_2,\cdots,k_M\}}= \mathbf{s}^{(k_1)}\circ \mathbf{s}^{(k_2)} \circ \dots \circ \mathbf{s}^{(k_M)} \eqno{(11)}$$ 
 for $2\le M\le N. $  
     In particular $\mathcal{T}^{(N)}=\mathbf{s}^{(1)}\circ \mathbf{s}^{(2)} \circ \dots \circ \mathbf{s}^{(N)}$ 
    holds. Here $\{k_1,k_2,\dots,k_M\}\subset \{1,2,\dots,N\}$, and $ \mathbf{s}^{(k)}$ is the Bloch vector of $k$th subsystem reduced density matrix.
 \\
 
 \textbf{ Proof} : Notice that Eq.(8) can be rewritten as

  $$\rho=\rho^{(1)} \otimes\rho^{(2)} \otimes \dots \otimes \rho^{(N)}+\fr{1}{d_1 d_2 \cdots d_N}\{\sum_{\{k_1,k_2\}}\sum_{\alpha_{k_1}\alpha_{k_2}}[t_{\alpha_{k_1}\alpha_{k_2}}-s_{\alpha_{k_1}} s_{\alpha_{k_2}}]\lambda^{(k_1)}_{\alpha_{k_1}} \lambda^{(k_2)}_{\alpha_{k_2}}+\cdots +$$
  $$+\cdots+\sum_{\alpha_{1}\alpha_{2}\cdots \alpha_{N}}[t_{\alpha_{1}\alpha_{2}\cdots \alpha_{N}}-s_{\alpha_{1}}s_{\alpha_{2}}\dots s_{\alpha_{N}}]\lambda^{(1)}_{\alpha_{1}} \lambda^{(2)}_{\alpha_{2}}\cdots \lambda^{(N)}_{\alpha_{N}}\}. \eqno{(12)} $$
   
   For full separability, the sum of all the  terms apart from the first term must vanish.
   Note that for every subsystem $k=1,2,\dots,N$ the set $\{I_d,\lambda_i; i=1,2,\dots,d_{k}^2-1\}$ forms an orthonormal Hilbert-Schmidt basis for the $k$th subsystem. Hence $ \lambda^{(k)}_{\alpha_{k}} ; \lambda^{(k_1)}_{\alpha_{k_1}} \lambda^{(k_2)}_{\alpha_{k_2}} \dots;\\
    \lambda^{(k_1)}_{\alpha_{k_1}} \lambda^{(k_2)}_{\alpha_{k_2}}\cdots \lambda^{(k_M)}_{\alpha_{k_M}};\dots ; \lambda^{(1)}_{\alpha_{1}} \lambda^{(2)}_{\alpha_{2}}\cdots \lambda^{(N)}_{\alpha_{N}} $ are the vectors belonging to the orthonormal product basis of the Hilbert-Schmidt space of the whole $N$-partite system. By orthonormality of the tensor product of $\lambda$'s occurring in different terms, the required sum will vanish if and only if coefficients of each term vanish separately,  that is if and only if\\
    
   $ t_{\alpha_{k_1}\alpha_{k_2}\cdots \alpha_{k_M}}=s_{\alpha_{k_1}}s_{\alpha_{k_2}}\dots s_{\alpha_{k_M}}; \; 2 \le M \le N,$\\  
    
that is,\\
    
$\mathcal{T}^{\{k_1,k_2,\cdots,k_M\}}=\mathbf{s}^{(k_1)}\circ \mathbf{s}^{(k_2)} \circ \dots \circ \mathbf{s}^{(k_M)}   \; ; 2 \le M \le N.$ \hfill $\square$\\

In fact, the condition (11) for all $N$ parts is enough to decide the separability of pure $N$-partite quantum states, as the following proposition shows.

\textbf{ Proposition 1a }: A  pure $N$-partite quantum state with Bloch representation (8) is fully separable (product state) if and only if   $$\mathcal{T}^{(N)}=\mathbf{s}^{(1)}\circ \mathbf{s}^{(2)} \circ \dots \circ \mathbf{s}^{(N)},$$ 
     where $ \mathbf{s}^{(k)}$ is the Bloch vector of $k$th subsystem reduced density matrix.\\
    
\textbf{ Proof} : Suppose $\rho$ is a product state $\rho= \rho_1 \otimes \rho_2 \otimes \cdots \otimes \rho_N.$ Then,
$$t_{\alpha_{1}\alpha_{2}\dots\alpha_{N}}=\fr{d_{1}d_{2}\dots d_{N}}{2^N} Tr[(\rho_1 \otimes \rho_2 \otimes \cdots\otimes \rho_N )(\lambda_{\alpha_{1}}\otimes \lambda_{\alpha_{2}}\otimes\cdots \otimes\lambda_{\alpha_{N}})]$$
$$=\fr{d_{1}d_{2}\dots d_{N}}{2^N} Tr[(\rho_1\lambda_{\alpha_{1}}) \otimes (\rho_2 \lambda_{\alpha_{2}})\otimes \cdots\otimes (\rho_N \lambda_{\alpha_{N}})]$$

$$=\fr{d_{1}d_{2}\dots d_{N}}{2^N} [Tr(\rho_1\lambda_{\alpha_{1}}) Tr(\rho_2 \lambda_{\alpha_{2}}) \cdots Tr( \rho_N \lambda_{\alpha_{N}})]$$
$$=[s_{\alpha_{1}}s_{\alpha_{2}}\cdots s_{\alpha_{N}}].$$

Suppose the condition holds, that is, $[s^{(1)}\circ s^{(2)}\circ \cdots \circ s^{(N)}]_{\alpha_{1}\alpha_{2}\dots\alpha_{N}}= t_{\alpha_{1}\alpha_{2}\dots\alpha_{N}}$. Then,

$$[s^{(1)}\circ s^{(2)}\circ \cdots \circ s^{(N)}]_{\alpha_{1}\alpha_{2}\dots\alpha_{N}}=\fr{d_{1}d_{2}\dots d_{N}}{2^N}[Tr(\rho_1\lambda_{\alpha_{1}}) Tr(\rho_2 \lambda_{\alpha_{2}}) \cdots Tr( \rho_N \lambda_{\alpha_{N}})]$$
$$=\fr{d_{1}d_{2}\dots d_{N}}{2^N} Tr[(\rho_1\lambda_{\alpha_{1}}) \otimes (\rho_2 \lambda_{\alpha_{2}})\otimes \cdots\otimes (\rho_N \lambda_{\alpha_{N}})]$$
$$=\fr{d_{1}d_{2}\dots d_{N}}{2^N} Tr[(\rho_1 \otimes \rho_2 \otimes \cdots\otimes \rho_N )(\lambda_{\alpha_{1}}\otimes \lambda_{\alpha_{2}}\otimes\cdots\otimes \lambda_{\alpha_{N}})]$$

$$=t_{\alpha_{1}\alpha_{2}\dots\alpha_{N}}= \fr{d_{1}d_{2}\dots d_{N}}{2^N}Tr[\rho(\lambda_{\alpha_{1}} \otimes \lambda_{\alpha_{2}} \otimes \cdots \otimes \lambda_{\alpha_{N}})].$$
The equality
$$Tr[(\rho_1 \otimes \rho_2 \otimes \cdots\otimes \rho_N )(\lambda_{\alpha_{1}}\otimes \lambda_{\alpha_{2}}\otimes \cdots \otimes \lambda_{\alpha_{N}})]= Tr[\rho(\lambda_{\alpha_{1}}\otimes \lambda_{\alpha_{2}} \otimes \cdots  \otimes \lambda_{\alpha_{N}})]$$
is satisfied for all elements in the orthonormal basis $\{\otimes_{k=1}^N \lambda_{\alpha_k}\},\; 0 \le \alpha_k \le d_k^2-1,\; (\alpha_k=0$ for $I_{d_k})$ where $\{\lambda_{\alpha_k}\}$ are the $d^2_k-1$ generators of $SU(d_k)$. This means that the joint probabilities obtained from the ensemble of measurements of $(\lambda_{\alpha_1} \cdots \lambda_{\alpha_N}) $ for states $\rho$ and $\rho=\rho_1 \otimes \rho_2 \otimes \cdots\otimes \rho_N$ are equal.
This implies $$\rho=\rho_1 \otimes \rho_2 \otimes \cdots\otimes \rho_N.$$ \hfill $\square$\\

Note that this criterion is easily amenable with experiments. In order to check it for an element of $\mathcal{T}^{(N)}$ we have to measure the corresponding generators on each subsystem and then check whether the product of the averages equals the average of the products.

Thus in order to check whether a given pure state is a product state we have to check whether $\mathcal{T}^{(N)}=\mathbf{s}^{(1)}\circ \mathbf{s}^{(2)} \circ \dots \circ \mathbf{s}^{(N)}$, where the Bloch vectors $\mathbf{s}^{(1)},  \mathbf{s}^{(2)}, \dots , \mathbf{s}^{(N)}$ can be constructed from the reduced density matrices $\rho_{1}, \rho_{2}, \cdots, \rho_{N}$ for subsystems $1,2,\cdots, N \; (s_{\alpha_{k}}=\fr{d_k}{2} Tr(\rho_{k} \la_{\alpha_{k}}),\; k \in \{1,2,\cdots,N\}$, see Eq.(9a)).
   
In the case of  mixed states we can characterize separability from the Bloch representation point of view as follows.
   
  {\it A $N$-partite quantum state with Bloch representation (8) is fully separable if and only if there exist vectors $u_w^{(k)} \in \mathbb{R}^{d_{k}^2-1}$ satisfying Eq.(5), and weights $p_w$  satisfying $0 \le p_w \le 1$ and $\sum_w p_w =1$ such that  $$ \mathcal{T}^{(N)}=\sum_w^R p_w \bigcirc_{k=1}^N u_w^{(k)}, \; \; \;  \mathbf{s}^{(k)}=\sum_w p_w u_w^{(k)} \eqno{(13a)}$$   and
  
 $$\mathcal{T}^{\{k_1,k_2,\cdots,k_M\}} = \sum_w^R p_w \bigcirc_{i=1}^M u_w^{(k_i)}\eqno{(13b)}$$ for $2\le M\le N $ ; for all subsets $\{k_1,k_2,\dots,k_M\}\subset \{1,2,\dots,N\},$} \\
 
 where $\mathbf{s}^{(k)}$  is the Bloch vector of the mixed state density matrix  for $k$th subsystem  and $u_w^{(k)}$ represent the Bloch vector of the pure state of the $k$th subsystem contributing to the $w$th term in Eq. (1).
 
 This follows from proposition 1 and Eq. (1). However, in view of proposition (1a), the necessary and sufficient condition is given by Eq.(13a), so that Eq.(13b) can be dropped. The above result can not be used directly, as it amounts to rewriting Werner's definition of separability in a different way.  
  However, it allows us to derive a necessary condition for separability for $N$-partite quantum states. 
  
  We need some concepts in multilinear algebra. Consider a tensor $\mathcal{T}^{(N)}\in \mathbb{R}^{I_1 \times I_2 \times \dots \times I_N}$ , where $I_k=d_k^2-1$. The $n$th matrix unfolding of $\mathcal{T}^{(N)}$ $(n=1,2,\cdots,N)$ [10] is a matrix  
 $T^{(N)}_{(n)}\in \mathbb{R}^{I_n \times (I_{n+1} I_{n+2}  \dots  I_N I_1 I_2 \dots I_{n-1})}$. $T^{(N)}_{(n)}$ contains the element $t_{i_1 i_2 \dots i_N} $ at the position with row index $i_n$  $(i_n=1,2,\cdots,I_n)$ and column index 
 
 $$(i_{n+1} - 1)I_{n+2}I_{n+3} \dots I_N I_1 I_2 \dots I_{n-1} + (i_{n+2} - 1)I_{n+3}I_{n+4}\dots I_N I_1 I_2 \dots I_{n-1} $$
 $$+ \dots +(i_N - 1)I_1 I_2 \dots I_{n-1} + (i_1 - 1)I_2I_3 \dots I_{n-1} + (i_2 - 1)I_3I_4 \dots I_{n-1} +\dots + i_{n-1}.$$
 
  For $n=1$, we take the last term $i_{n-1}=i_{0}=i_N$. This ordering is called backward cyclic [16].  To facilitate understanding, put $N$ points on a circle and label them successively by  $i_1,i_2,\cdots,i_N.$ The consecutive terms in the expression for the column index in $T^{(N)}_{(n)}$ corresponding to $t_{i_1,i_2,\cdots,i_N}$ become quite apparent using this circle.
  
  For  $\mathcal{T}^{(3)} \in \mathbb{R}^{I_1 \times I_2 \times I_3}$ the matrix unfolding $T^{(3)}_{(1)}$ contains the elements $t_{i_1i_2i_3} \; (i_k=1,2,\cdots,I_k ; \; k=1,2,3)$ at the position with row number $i_1$ and column number equal to $(i_2-1)I_3+i_3$, $T^{(3)}_{(2)}$ contains $t_{i_1i_2i_3}$ at the position with row number $i_2$ and column number equal to $(i_3-1)I_1+i_1$ and $T^{(3)}_{(3)}$ contains $t_{i_1i_2i_3}$ at the position with row number $i_3$ and column number equal to $(i_1-1)I_2+i_2.$

 As an example [25], define a tensor $\mathcal{T}^{(3)} \in \mathbb{R}^{3\times 2 \times 3}$, by $t_{111} = t_{112} = t_{211} =-t_{212} = 1$, $t_{213} = t_{311} = t_{313} = t_{121} = t_{122} = t_{221} = -t_{222} = 2$, $t_{223} = t_{321} = t_{323} = 4$,
$t_{113} = t_{312} = t_{123} = t_{322} = 0$. The matrix unfolding $T^{(3)}_{(1)}$ is given by

\begin{displaymath}
T^{(3)}_{(1)} =
\left(\begin{array}{ccc|ccc}
1& 1 & 0 & 2 & 2 & 0 \\
1 & -1 & 2 & 2 & -2 & 4\\
2 & 0 & 2 & 4 & 0 & 4
\end{array}\right).
\end{displaymath}

 Note that there are $N$ possible matrix unfoldings of $\mathcal{T}^{(N)}$. The matrix unfolding is called the matrization of the tensor [10,17]. We can now define the Ky Fan norm of the tensor $\mathcal{T}^{(N)}$ (of order $N$)   over $N$ matrix unfoldings of a tensor, as $$||\mathcal{T}^{(N)}||_{KF} = max\{||T^{(N)}_{(n)}||_{KF}\}, \; n=1,\dots,N ; \eqno{(14)}$$ 
 where $||T^{(N)}_{(n)}||_{KF}$ is the  Ky Fan norm of matrix $T^{(N)}_{(n)}$ defined as the sum of singular values of $T^{(N)}_{(n)}$ [26]. It is straightforward to check that $||\mathcal{T}^{(N)}||_{KF}$ defined in (14) satisfies all the conditions of a norm and is also unitarily invariant [9,26].\\
 
  The tensors in Eq.(13a) are called  Kruskal tensors with the restriction $0 \le p_w\le 1,\;  \sum_w p_w=1$ [14,16]. We are interested in finding the matrix unfoldings and Ky Fan norms of $\mathcal{T}^{(N)}$ occurring in Eq.(13a).    The  $k$th matrix unfolding for Kruskal tensor is [17]
 
 $$T_{(k)}^{(N)} = U^{(k)} \Sigma (U^{(N)}\odot U^{(N-1)}\odot \dots \odot U^{(k+1)}\odot U^{(k-1)} \odot \dots \odot U^{(1)})^T . \eqno{(15)}$$
 
  Here  $U^{(k)}=[\mathbf{u}_1^{(k)}  \mathbf{u}_2^{(k)} \dots \mathbf{u}_R^{(k)}] \in  \mathbb{R}^{I_{k} \times R}; k =1,2,\dots N$ and  $R$ is the rank of Kruskal tensor [14,12,17], i.e. the number of terms in Eq.(13a). $\mathbf{u}_i^{(k)}$ is a vector in  $\mathbb{R}^{I_{k}}$ and is the $i$th column vector in the matrix $U^{(k)}.$ $\Sigma$ is the $R \times R$ diagonal matrix,  $\Sigma =$diag$[p_1 \dots p_R]$. The symbol $\odot$ denotes the Khatri-Rao product of matrices [17] $U \in  \mathbb{R}^{I\times R}$   and $V \in \mathbb{R}^{J\times R}$ defined as $U\odot V =[ \mathbf{u}_1\otimes \mathbf{v}_1 \; \mathbf{u}_2 \otimes \mathbf{v}_2 \; \dots \; \mathbf{u}_R \otimes \mathbf{v}_R] \in \mathbb{R}^{IJ\times R}$ where $\mathbf{u}_i$ and $\mathbf{v}_i ,\; i=1,2,\dots R $ are column vectors of matrices $U$ and $V$ respectively. Eq.(15) can be rewritten as 
    
 $$T_{(k)}^{(N)} = U^{(k)} \Sigma [\mathbf{v}_1^{(\bar{k})} \mathbf{v}_2^{(\bar{k})}\dots \mathbf{v}_R^{(\bar{k})}]^T = U^{(k)} \Sigma V^{(\bar{k})^T}  \eqno{(16)}$$ 
 where $\mathbf{v}_i^{(\bar{k})} ; i=1,2,\dots , R$ are the column vectors of the matrix \\ 
 $V^{(\bar{k})^T} \in \mathbb{R}^{I_{N} I_{N-1} I_{N-2}  \dots  I_{k+1}I_{k-1}\dots I_{1} \times R}$   and  $ \mathbf{v}_i^{\bar{(k)}} = \mathbf{u}_i^{(N)} \otimes \mathbf{u}_i^{(N-1)}\otimes \mathbf{u}_i^{(N-2)}\otimes \dots \otimes \mathbf{u}_i^{(k+1)} \otimes \mathbf{u}_i^{(k-1)} \otimes \dots \otimes \mathbf{u}_i^{(1)}.$ 
  Using Eq.(16) we can write $T_{(k)}^{(N)}$ as  $$T_{(k)}^{(N)}=\sum_{w=1}^R p_w \mathbf{u}_w^{(k)} \mathbf{v}_w^{(\bar{k})^T} ;\; \;k=1,2,\dots ,N.  \eqno{(17)}$$
 
  \textbf{ Theorem 1 }: If a $N$-partite quantum state of $d_1 d_2 \dots d_N$ dimension with Bloch representation (8) is fully  separable, then $$||\mathcal{T}^{(N)}||_{KF} \le \sqrt{\fr{1}{2^N} \Pi_{k=1}^N d_{k}(d_{k}-1)}. \eqno{(18)}$$

  \textbf{ Proof} : If the state $\rho$ is separable then $\mathcal{T}^{(N)}$ has to admit a decomposition of the form Eqs.(13) with $||\mathbf{u}_w^{(k)}||_2= \sqrt{\fr{d_{k}(d_{k}-1)}{2}}, k = 1,2,\dots,N.$
   From definition of KF norm of tensors, Eq.(14),

    $$||\mathcal{T}^{(N)}||_{KF} =max \{||T_{(k)}^{(N)}||_{KF}\}\; ;\; k=1,\dots, N.$$
   From Eq.(17), 
    
    $$||\mathcal{T}^{(N)}||_{KF} =max \{||\sum_w p_w \mathbf{u}_w^{(k)} \mathbf{v}_w^{(\bar{k})^T} ||_{KF}\} \; ;\; k=1,\dots, N$$
    $$\le max \{\sum_w p_w ||\mathbf{u}_w^{(k)} \mathbf{v}_w^{(\bar{k})^T} ||_{KF}\}  
     =max\{ \sum_w p_w \sqrt{{\fr{1}{2^N}} \Pi_k^N d_{k}(d_{k}-1)}|| \mathbf{\tilde{u}}_w^{(k)} \mathbf{\tilde{v}}_w^{(\bar{k})^T} ||_{KF}\}$$
     
       where $\mathbf{\tilde{u}}_w^{(k)}, \mathbf{\tilde{v}}_w^{(\bar{k})}$ are unit vectors in 
   $ \mathbb{R}^{d_{k}^2-1}$ and $\mathbb{R}^{d_{N}^2-1}\otimes \mathbb{R}^{d_{N-1}^2-1}\otimes \dots \otimes \mathbb{R}^{d_{k+1}^2-1}\otimes  \mathbb{R}^{d_{k-1}^2-1}\otimes \dots \otimes \mathbb{R}^{d_1^2-1}$ respectively, so that  
 $|| \mathbf{\tilde{u}}_w^{(k)} \mathbf{\tilde{v}}_w^{(\bar{k})^T} ||_{KF}=1$ for all $k=1,2,\dots,N$. Using $\sum_w p_w=1$ we get $||\mathcal{T}^{(N)}||_{KF} \le \sqrt{\fr{1}{2^N} \Pi_{k=1}^N d_{k}(d_{k}-1)}$.\hfill $\square$\\
 
 For a subsystem we get,
 
 \textbf{ Corollary 1 }: If the reduced density matrix of a subsystem consisting of $M$ out of $N$ parts is separable then 
 $||\mathcal{T}^{\{k_1,k_2,\cdots,k_M \}}||_{KF} \le \sqrt{\fr{1}{2^M} \Pi_{k=1}^M d_{k}(d_{k}-1)}$.
 
  The negation of the above condition, that is, $||\mathcal{T}^{(N)}||_{KF} > \sqrt{\fr{1}{2^N} \Pi_{k=1}^N d_{k}(d_{k}-1)}$,  is a sufficient condition of entanglement of $N$-partite quantum state. This leads to a hierarchy of inseparability conditions  which test entanglement in all the subsystems. For $N=2$ the condition $||\mathcal{T}^{(N)}||_{KF} \le \sqrt{\fr{1}{2^2} d_1(d_1-1)d_2(d_2-1)} $ has been shown in Ref. [9], to be a sufficient condition for entanglement associated with any bipartite density matrix. Note that for $N$-qubits, ${d_i=2,\; i=1,2,\dots,N}$, the above criterion  becomes, for a separable state, $||\mathcal{T}^{(N)}||_{KF} \le 1$.
   
  Consider a $N$ qudit system $\mathcal{H}_s=\otimes_{k=1}^N\mathcal{H}^{d}_k$ in a state $\rho$, supported in the symmetric subspace of $\mathcal{H}_s$. It is straightforward to see that all the tensors in the Bloch representation of $\rho$ are supersymmetric, that is (see Eqs.(8) and (9)),  
$ t_{\alpha_{k_1}\alpha_{k_2} \cdots \alpha_{k_M}}= t_{P(\alpha_{k_1})P(\alpha_{k_2}) \cdots P(\alpha_{k_M})}, \; 2 \le M \le N,$  where  $P$ is any permutation over indices  $\{\alpha_{k_1},\alpha_{k_2},\cdots ,\alpha_{k_M}\}$. We have, neglecting the constant multipliers,

\benrr
t_{\alpha_{k_1}\alpha_{k_2} \cdots \alpha_{k_M}}&=& Tr[\rho_{k_1k_2\cdots k_M}  \lambda_{\alpha_{k_1}}\otimes\lambda_{\alpha_{k_2}}\otimes\dots \otimes\lambda_{\alpha_{k_M}}]\\
&=& Tr[\rho_{k_1k_2\cdots k_M}P P^T \lambda_{\alpha_{k_1}}\otimes\lambda_{\alpha_{k_2}}\otimes\dots \otimes\lambda_{\alpha_{k_M}}PP^T]\\
 &=& Tr[(P^T\rho_{k_1k_2\cdots k_M}P) (P^T \lambda_{\alpha_{k_1}}\otimes\lambda_{\alpha_{k_2}}\otimes\dots \otimes\lambda_{\alpha_{k_M}}P)]\\
&= & Tr[\rho_{k_1k_2\cdots k_M} ( \lambda_{P(\alpha_{k_1})}\otimes\lambda_{P(\alpha_{k_2})}\otimes\dots \otimes\lambda_{P(\alpha_{k_M})})]\\
 &=& t _{P(\alpha_{k_1})P(\alpha_{k_2}) \cdots P(\alpha_{k_M})}
\eenrr
   
  where $P$ is the appropriate permutation matrix permuting the $\la$ matrices within the tensor product [26], $P^T$ being the transpose of $P$ satisfying $P^T=P^{-1}$. In particular $\mathcal{T}^{(N)}$ is supersymmetric. All matrix unfoldings of a supersymmetric tensor have the same set of singular values [10] and hence the same KF norm. Thus, for a $N$-qudit system in a state supported in the symmetric subspace, it is enough to calculate the KF norm for any one of the $N$ matrix unfoldings to get  $\max \{||T_{(k)}^{(N)}||_{KF}\}.$
 

 \section{ A Sufficient Condition for Separability of a 3-Partite Quantum State}
 
 Consider the Bloch representation of a tripartite state $\rho$ acting on $\mathcal{H}^{d_1}\otimes \mathcal{H}^{d_2} \otimes \mathcal{H}^{d_3}$,\; $d_1 \le d_2 \le d_3$.
  
 $$\rho = \fr{1}{d_1 d_2 d_3}(\otimes_{k=1}^3 I_{d_k}+\sum_{\alpha_1} r_{\alpha_1} \lambda^{(1)}_{\alpha_1}+\sum_{\alpha_2} s_{\alpha_2} \lambda^{(2)}_{\alpha_2}+\sum_{\alpha_3} q_{\alpha_3} \lambda^{(3)}_{\alpha_3}+\sum_{\alpha_1\alpha_2}t_{\alpha_1\alpha_2}  \lambda^{(1)}_{\alpha_1} \lambda^{(2)}_{\alpha_2}$$
 
 $$+\sum_{\alpha_1\alpha_3}t_{\alpha_1\alpha_3}  \lambda^{(1)}_{\alpha_1} \lambda^{(3)}_{\alpha_3}+\sum_{\alpha_2\alpha_3}t_{\alpha_2\alpha_3}  \lambda^{(2)}_{\alpha_2} \lambda^{(3)}_{\alpha_3} +\sum_{\alpha_1\alpha_2\alpha_3}t_{\alpha_1\alpha_2\alpha_3}  \lambda^{(1)}_{\alpha_1} \lambda^{(2)}_{\alpha_2}\lambda^{(3)}_{\alpha_3},\eqno{(19a)}$$
 
 where $\mathbf{r},\; \mathbf{s}$ and $\mathbf{q}$ are the Bloch vectors of three subsystems respectively , $T^{\{\mu,\nu\}}=[t_{\alpha_{\mu}\alpha_{\nu}}]$ the correlation matrix between the subsystems $\mu, \nu;  \{\mu,\nu\} \subset \{1,2,3\}$ and $\mathcal{T}^{(3)}=[ t_{\alpha_1\alpha_2\alpha_3}]$ the correlation tensor among three subsystems. Before stating proposition 2, we need the following definition and result.\\

Kruskal decomposition of a tensor $\mathcal{T}^{(N)}$ 

$$\mathcal{T}^{(N)}=\sum_{j=1}^R \xi_j \mathbf{u}_j^{(1)}\circ\mathbf{u}_j^{(2)}\circ\cdots \circ\mathbf{u}_j^{(N)}$$

is called completely orthogonal if $\lan u_k^{(i)},u_l^{(i)}\ran = \delta_{kl},\;i=1,2,\cdots,N;\;k,l=1,2,\cdots,R$ [13], where $\lan,\ran$ denotes the scalar product of two vectors.
If $\mathcal{T}^{(N)}$  has completely orthogonal Kruskal decomposition, then it is straightforward to show that $$||\mathcal{T}^{(N)}||_{KF}=\sum_{j=1}^R \xi_j,\eqno{(20)}$$
where $R$ is the rank of $\mathcal{T}^{(N)}$  and $\xi_j,\;j=1,2,\cdots,R$ are the coefficients occurring in the completely orthogonal Kruskal decomposition of $\mathcal{T}^{(N)}$. In the proof of proposition 2, we assume that completely orthogonal Kruskal decomposition of $\mathcal{T}^{(k)},\; k>2$ is available. A completely orthogonal Kruskal decomposition may not be available for an arbitrary tensor [13].  The general conditions under which the completely orthogonal Kruskal decomposition is possible is an open problem. We conjecture that completely orthogonal kruskal decomposition is available for all tensors in the Bloch representation of a quantum state, but we do not have a proof.
As it stands, this issue has to be settled case by case.\\

\textbf{ Proposition 2 }: If a tripartite state $\rho$ acting on  $\mathcal{H}^{d_1}\otimes \mathcal{H}^{d_2} \otimes \mathcal{H}^{d_3}$,\; $d_1 \le d_2 \le d_3$, with Bloch representation (19a), where $\mathcal{T}^{(3)}$ has the completely orthogonal Kruskal decomposition, satisfies 

$$\sq{\fr{2(d_1-1)}{d_1}} ||\mathbf{r}||_2+\sq{\fr{2(d_2-1)}{d_2}} ||\mathbf{s}||_2+\sq{\fr{2(d_3-1)}{d_3}} ||\mathbf{q}||_2+\sum_{\{\mu,\nu\}}\sq{\fr{4(d_{\mu}-1)(d_{\nu}-1)}{d_{\mu}d_{\nu}}} ||T^{\{\mu,\nu\}}||_{KF}$$

$$+ \sq{\fr{8(d_1-1)(d_2-1)(d_3-1)}{d_1d_2d_3}} ||\mathcal{T}||_{KF} \le 1,  \eqno{(21)}$$
 
 then $\rho$ is separable.\\

\textbf{ Proof} : The idea of the proof is as follows.\\
 
(i) We first decompose all the tensors in the Bloch representation of $\rho$ as the completely orthogonal Kruskal decomposition in terms of the outer products of the vectors in the Bloch spaces of the subsystems (coherence vectors).

(ii)We prove that we can decompose $\rho$ using the Kruskal decompositions described in (i) above, as the linear combination of separable density matrices, which is a convex combination if the coefficient of identity is positive. This condition is the same as the condition stated in the proposition.

Let $T^{\{\mu,\nu\}} ;\{\mu,\nu\} \subset \{1,2,3\}$ in Eq.(19a) have singular value decomposition $T^{\{\mu,\nu\}} = \sum_i \si_i \mathbf{a}^{(\mu)}_i ({\mathbf{a}^{(\nu)}_i})^T \; ; \; $ with $||\mathbf{a}_i^{(\mu)}||_2=||\mathbf{a}_i^{(\nu)}||_2=1$ , for $\{\mu, \nu \}\subset \{1,2,3\}$ and  let $\mathcal{T}$ in Eq. (19a) have the completely orthogonal Kruskal decomposition $\mathcal{T}= \sum_j \xi_j \mathbf{u}_j\circ \mathbf{v}_j \circ \mathbf{w}_j$ [17,14,27] with $||\mathbf{u}_j||_2 = ||\mathbf{v}_j||_2=||\mathbf{w}_j||_2=1$. We define\\

 $\mathbf{\tilde{a}}^{(\mu)}_i=\sq{\fr{d_{\mu}}{2(d_{\mu}-1)}}\;\mathbf{a}^{(\mu)}_i$  \; , \;  $\mu  \in \{1,2,3\}$\\
 
 so that we can rewrite  $$T^{\{\mu,\nu\}} = \sq{\fr{4(d_{\mu}-1)(d_{\nu}-1)}{d_{\mu}d_{\nu}}} \sum_i \si_i \mathbf{\tilde{a}}^{\mu}_i (\mathbf{\tilde{a}}^{\nu}_i)^T \eqno{(22a)}$$.
 
  Similarly, we define \\
  
 $\mathbf{\tilde{u}}_j=\sq{\fr{d_1}{2(d_1-1)}} \; \mathbf{u}_j$\; ; \; $\mathbf{\tilde{v}}_j=\sq{\fr{d_2}{2(d_2-1)}} \; \mathbf{v}_j$\; ; \;  $\mathbf{\tilde{w}}_j=\sq{\fr{d_3}{2(d_3-1)}} \; \mathbf{w}_j$,   so that we can write $$\mathcal{T}=\sq{\fr{8(d_1-1)(d_2-1)(d_3-1)}{d_1d_2d_3}} \sum_j \xi_j \mathbf{\tilde{u}}_j\circ \mathbf{\tilde{v}}_j \circ \mathbf{\tilde{w}}_j \eqno{(22b)}$$
 
 If we substitute Eqs.(22a) and (22b) in $\rho$ Eq.(19a), we get
 
 $$\rho = \fr{1}{d_1 d_2 d_3}(\otimes_{k=1}^3 I_{d_k}+\sum_{\alpha_1} r_{\alpha_1} \lambda^{(1)}_{\alpha_1}+\sum_{\alpha_2} s_{\alpha_2} \lambda^{(2)}_{\alpha_2}+\sum_{\alpha_3} q_{\alpha_3} \lambda^{(3)}_{\alpha_3}$$
$$+\sum_{\{\mu,\nu\}}\sum_{\alpha_{\mu}\alpha_{\nu}}\sq{\fr{4(d_{\mu}-1)(d_{\nu}-1)}{d_{\mu}d_{\nu}}}\sum_i\si_i(\mathbf{\tilde{a}}^{(\mu)}_{i})_{\alpha_{\mu}}(\mathbf{\tilde{a}}^{(\nu)}_i)_{\alpha_{\nu}}\lambda^{(\mu)}_{\alpha_{\mu}} \lambda^{(\nu)}_{\alpha_{\nu}}$$
$$+\sq{\fr{8(d_1-1)(d_2-1)(d_3-1)}{d_1d_2d_3}}\sum_{\alpha_1\alpha_2\alpha_3}\sum_j\xi_j(\mathbf{\tilde{u}}_j)_{\alpha_1}(\mathbf{\tilde{v}}_j)_{\alpha_2}(\mathbf{\tilde{w}}_j)_{\alpha_3}  \lambda^{(1)}_{\alpha_1} \lambda^{(2)}_{\alpha_2}\lambda^{(3)}_{\alpha_3} \eqno{(19b)}$$
 
 The coherence vectors $\mathbf{\tilde{a}}^{(\mu)}_i$ occur in $D_r(\mathbb{R}^{d_{\mu}^2-1})$, $\mathbf{\tilde{a}}^{(\nu)}_i$ occur in $D_r(\mathbb{R}^{d_{\nu}^2-1})$, $\mathbf{\tilde{u}}_j$ occur in $D_r(\mathbb{R}^{d_1^2-1})$, $\mathbf{\tilde{v}}_j$ occur in $D_r(\mathbb{R}^{d_2^2-1})$ and $\mathbf{\tilde{w}}_j$ occur in $D_r(\mathbb{R}^{d_3^2-1})$ (see Eq.(6)), so that they correspond to Bloch vectors.
 
 We can decompose $\rho$ Eq.(19b) as the following convex combination of the density matrices 
  
  $ \rho_j \; , \; \rho'_j \; ,\; \rho''_j \;,\; \rho'''_j$ ; $\varrho_i \;,\; \varrho'_i\;,\; \tau_i\;,\;\tau'_i\;,\; \pi_i\;,\;\pi'_i$ ; $\rho_r \;,\; \rho_s \; ,\; \rho_q$ and $\fr{1}{d_1d_2d_3} I_{d_1d_2d_3}$;
  
  $$\rho=\sum_j\sq{\fr{8(d_1-1)(d_2-1)(d_3-1)}{d_1d_2d_3}} \fr{\xi_j}{4}(\rho_j +\rho'_j +\rho''_j + \rho'''_j)+\sum_i\sq{\fr{4(d_1-1)(d_2-1)}{d_1d_2}} \fr{\si_i}{2}(\varrho_i+ \varrho'_i)$$
  $$+\sum_i\sq{\fr{4(d_1-1)(d_3-1)}{d_1d_3}} \fr{\si'_i}{2}(\tau_i+ \tau'_i)+\sum_i\sq{\fr{4(d_2-1)(d_3-1)}{d_2d_3}} \fr{\si''_i}{2}(\pi_i+ \pi'_i)$$
  $$+\sq{\fr{2(d_1-1)}{d_1}}||\mathbf{r}||_2 \rho_r+\sq{\fr{2(d_2-1)}{d_2}}||\mathbf{s}||_2 \rho_s+\sq{\fr{2(d_3-1)}{d_3}}||\mathbf{q}||_2 \rho_q+(1-\sq{\fr{2(d_1-1)}{d_1}} ||\mathbf{r}||_2$$
$$-\sq{\fr{2(d_2-1)}{d_2}}||\mathbf{s}||_2-\sq{\fr{2(d_3-1)}{d_3}}||\mathbf{q}||_2-\sum_{\{\mu,\nu\}}\sq{\fr{4(d_{\mu}-1)(d_{\nu}-1)}{d_{\mu}d_{\nu}}} ||T^{\{\mu,\nu\}}||_{KF} $$

  $$- \sq{\fr{8(d_1-1)(d_2-1)(d_3-1)}{d_1d_2d_3}} ||\mathcal{T}||_{KF})\fr{I_{d_1d_2d_3}}{d_1d_2d_3}. \eqno{(23)}$$
  
   where $ \rho_j$ in Bloch representation is 
   
   $$\rho_j= \fr{1}{d_1 d_2 d_3}\Big(\otimes_{k=1}^3 I_{d_k}+\sum_{\alpha_1} (\mathbf{\tilde{u}}_j)_{\alpha_1} \lambda^{(1)}_{\alpha_1}+\sum_{\alpha_2} (\mathbf{\tilde{v}}_j)_{\alpha_2} \lambda^{(2)}_{\alpha_2}+\sum_{\alpha_3} (\mathbf{\tilde{w}}_j)_{\alpha_3} \lambda^{(3)}_{\alpha_3}$$
$$+\sum_{\alpha_1\alpha_2}(\mathbf{\tilde{u}}_j)_{\alpha_1}(\mathbf{\tilde{v}}_j)_{\alpha_2}\lambda^{(1)}_{\alpha_{1}} \lambda^{(2)}_{\alpha_{2}}+\sum_{\alpha_1\alpha_3}(\mathbf{\tilde{u}}_j)_{\alpha_1}(\mathbf{\tilde{w}}_j)_{\alpha_3}\lambda^{(1)}_{\alpha_{1}} \lambda^{(3)}_{\alpha_{3}}+\sum_{\alpha_2\alpha_3}(\mathbf{\tilde{v}}_j)_{\alpha_2}(\mathbf{\tilde{w}}_j)_{\alpha_3}\lambda^{(2)}_{\alpha_{2}} \lambda^{(3)}_{\alpha_{3}}$$
$$+\sum_{\alpha_1\alpha_2\alpha_3}(\mathbf{\tilde{u}}_j)_{\alpha_1}(\mathbf{\tilde{v}}_j)_{\alpha_2}(\mathbf{\tilde{w}}_j)_{\alpha_3}  \lambda^{(1)}_{\alpha_1} \lambda^{(2)}_{\alpha_2}\lambda^{(3)}_{\alpha_3} \Big)$$
 
 $$= \fr{1}{d_1 d_2 d_3}(I_{d_1}+\sum_{\alpha_1}(\mathbf{\tilde{u}}_j)_{\alpha_1}\lambda^{(1)}_{\alpha_1})\otimes (I_{d_2}+\sum_{\alpha_2}(\mathbf{\tilde{v}}_j)_{\alpha_2}\lambda^{(2)}_{\alpha_2})\otimes(I_{d_3}+\sum_{\alpha_3}(\mathbf{\tilde{w}}_j)_{\alpha_3}\lambda^{(3)}_{\alpha_3}). \eqno{(24)}$$
  
  Note that $||\mathcal{T}||_{KF}$ in Eq.(23) is defined via Eq.(20), which is based on completely orthogonal Kruskal decomposition of  $\mathcal{T}.$
  
  The Bloch vectors, correlation matrices and correlation tensors of the density matrices $ \rho_j \; , \; \rho'_j \; ,\; \rho''_j \;,\; \rho'''_j$ ; $\varrho_i \;,\; \varrho'_i\;,\; \tau_i\;,\;\tau'_i\;,\; \pi_i\;,\;\pi'_i$ ; $\rho_r \;,\; \rho_s \; ,\; \rho_q$ are\\
  
For $\rho_j$,\\
  
  $\mathbf{r}_j= \mathbf{\tilde{u}}_j \; ,\; \mathbf{s}_j=\mathbf{\tilde{v}}_j \; ,\; \mathbf{q}_j=\mathbf{\tilde{w}}_j \; ,\;  T^{\{1,2\}}_j=\mathbf{\tilde{u}}_j {\mathbf{\tilde{v}}_j}^T \;,\; T^{\{1,3\}}_j=\mathbf{\tilde{u}}_j {\mathbf{\tilde{w}}_j}^T \; ,\; T^{\{2,3\}}_j=\mathbf{\tilde{v}}_j {\mathbf{\tilde{w}}_j}^T$
  
   $\mathcal{T}_j=\mathbf{\tilde{u}}_j\circ \mathbf{\tilde{v}}_j\circ \mathbf{\tilde{w}}_j$.\\
  
  For $\rho'_j$,\\
  
$\mathbf{r'}_j= \mathbf{\tilde{u}}_j \; ,\; \mathbf{s'}_j=-\mathbf{\tilde{v}}_j \; ,\;\mathbf{q'}_j=-\mathbf{\tilde{w}}_j \; ,\;  T'^{\{1,2\}}_j=-\mathbf{\tilde{u}}_j {\mathbf{\tilde{v}}_j}^T \;,\; T'^{\{1,3\}}_j=-\mathbf{\tilde{u}}_j {\mathbf{\tilde{w}}_j}^T $ 

$T'^{\{2,3\}}_j=\mathbf{\tilde{v}}_j {\mathbf{\tilde{w}}_j}^T\;,\; \mathcal{T'}_j=\mathbf{\tilde{u}}_j\circ \mathbf{\tilde{v}}_j\circ \mathbf{\tilde{w}}_j$.\\
 
 For $\rho''_j$,\\
 
 $\mathbf{r''}_j= -\mathbf{\tilde{u}}_j \; ,\; \mathbf{s''}_j=\mathbf{\tilde{v}}_j \; ,\; \mathbf{q''}_j=-\mathbf{\tilde{w}}_j \; ,\;  T''^{\{1,2\}}_j=-\mathbf{\tilde{u}}_j {\mathbf{\tilde{v}}_j}^T \;,\; T''^{\{1,3\}}_j=\mathbf{\tilde{u}}_j {\mathbf{\tilde{w}}_j}^T$
 
 $ T''^{\{2,3\}}_j=-\mathbf{\tilde{v}}_j {\mathbf{\tilde{w}}_j}^T\;,\; \mathcal{T''}_j=\mathbf{\tilde{u}}_j\circ \mathbf{\tilde{v}}_j\circ \mathbf{\tilde{w}}_j$.\\

For $\rho'''_j$,\\

$\mathbf{r'''}_j=- \mathbf{\tilde{u}}_j \; ,\; \mathbf{s'''}_j=-\mathbf{\tilde{v}}_j \; ,\;\mathbf{q'''}_j=\mathbf{\tilde{w}}_j \; ,\;  T'''^{\{1,2\}}_j=\mathbf{\tilde{u}}_j {\mathbf{\tilde{v}}_j}^T \;,\; T'''^{\{1,3\}}_j=-\mathbf{\tilde{u}}_j {\mathbf{\tilde{w}}_j}^T$

$ T'''^{\{2,3\}}_j=-\mathbf{\tilde{v}}_j {\mathbf{\tilde{w}}_j}^T\;,\; \mathcal{T'''}_j=\mathbf{\tilde{u}}_j\circ \mathbf{\tilde{v}}_j\circ \mathbf{\tilde{w}}_j$.\\

For $\varrho_i$,\\

$\mathbf{r}^{\varrho}_i=\mathbf{\tilde{a}}_i^{(1)} \;,\; \mathbf{s}^{\varrho}_i=\mathbf{\tilde{a}}_i^{(2)} \; ,\; \mathbf{q}^{\varrho}_i=0\; ,\; T^{\varrho\{1,2\}}_i=\mathbf{\tilde{a}}_i^{(1)} {\mathbf{\tilde{a}}_i^{(2)T}} \;,\; T^{\varrho\{1,3\}}_i=0$

$ T^{\varrho\{2,3\}}_i=0\;,\; \mathcal{T}^{\varrho}_i=0$.\\

For $\varrho'_i$,\\

$\mathbf{r}^{\varrho'}_i=-\mathbf{\tilde{a}}_i^{(1)} \;,\; \mathbf{s}^{\varrho'}_i=-\mathbf{\tilde{a}}_i^{(2)} \; ,\; \mathbf{q}^{\varrho'}_i=0\; ,\; T^{\varrho'\{1,2\}}_i=\mathbf{\tilde{a}}_i^{(1)} {\mathbf{\tilde{a}}_i^{(2)T}}$

$T^{\varrho'\{1,3\}}_i=0 \; ,\; T^{\varrho'\{2,3\}}_i=0\;,\; \mathcal{T}^{\varrho'}_i=0$.\\

For $\tau_i$,\\

$\mathbf{r}^{\tau}_i=\mathbf{\tilde{a}}_i^{(1)} \;,\; \mathbf{s}^{\tau}_i=0 \; ,\; \mathbf{q}^{\tau}_i=\mathbf{\tilde{a}}_i^{(3)}\; ,\; T^{\tau^\{1,2\}}_i=0\;,\; T^{\tau\{1,3\}}_i=\mathbf{\tilde{a}}_i^{(1)} {\mathbf{\tilde{a}}_i^{(3)T}}$

$ T^{\tau\{2,3\}}_i=0\;,\; \mathcal{T}^{\tau}_i=0$.\\

For $\tau'_i$\\

$\mathbf{r}^{\tau'}_i=-\mathbf{\tilde{a}}_i^{(1)} \;,\;\mathbf{s}^{\tau'}_i=0 \; ,\; \mathbf{q}^{\tau'}_i=-\mathbf{\tilde{a}}_i^{(3)}\; ,\; T^{\tau'\{1,2\}}_i=0\;,\; T^{\tau'\{1,3\}}_i=\mathbf{\tilde{a}}_i^{(1)} {\mathbf{\tilde{a}}_i^{(3)T}}$

$ T^{\tau'\{2,3\}}_i=0\;,\; \mathcal{T}^{\tau'}_i=0$.\\

For $\pi$,\\

$\mathbf{r}^{\pi}_i=0 \;,\; \mathbf{s}^{\pi}_i=\mathbf{\tilde{a}}_i^{(2)} \; ,\; \mathbf{q}^{\pi}_i=\mathbf{\tilde{a}}_i^{(3)}\; ,\; T^{\pi}{\{1,2\}}_i=0\;,\; T^{\pi\{1,3\}}_i=0$

$ T^{\pi\{2,3\}}_i=\mathbf{\tilde{a}}_i^{(2)} {\mathbf{\tilde{a}}_i^{(3)T}}\;,\; \mathcal{T}^{\pi}_i=0$.\\

For $\pi'$,\\

$\mathbf{r}^{\pi'}_i=0 \;,\; \mathbf{s}^{\pi'}_i=-\mathbf{\tilde{a}}_i^{(2)} \; ,\;\mathbf{q}^{\pi'}_i=-\mathbf{\tilde{a}}_i^{(3)}\; ,\; T^{\pi'\{1,2\}}_i=0\;,\; T^{\pi'\{1,3\}}_i=0$

$T^{\pi'\{2,3\}}_i=\mathbf{\tilde{a}}_i^{(2)} {\mathbf{\tilde{a}}_i^{(3)T}}\;,\; \mathcal{T}^{\pi'}_i=0$.\\

For $\rho_r$,\\

$\mathbf{r}_r=\sq{\fr{d_1}{2(d_1-1)}} \fr{\mathbf{r}}{||\mathbf{r}||_2} \;,\; \mathbf{s}_r=0 \;, \; \mathbf{q}_r=0  \;,\; T^{\{\mu,\nu\}}_r=0\;;\; \forall \{\mu,\nu\} \subset \{1,2,3\} \;,\; \mathcal{T}_r=0$.\\

For $\rho_s$,\\

$\mathbf{r}_s=0 \;,\; \mathbf{s}_s= \sq{\fr{d_2}{2(d_2-1)}} \fr{\mathbf{s}}{||\mathbf{s}||_2} \;,\; \mathbf{q}_s=0 \;,\; T^{\{\mu,\nu\}}_s=0 \;;\; \forall \{\mu,\nu\} \subset \{1,2,3\} \;,\; \mathcal{T}_s=0$.\\

For $\rho_q$,\\

$\mathbf{r}_q=0 \;,\;\mathbf{s}_q=0 \;,\; \mathbf{q}_q=\sq{\fr{d_3}{2(d_3-1)}} \fr{\mathbf{q}}{||\mathbf{q}||_2} \;,\; T^{\{\mu,\nu\}}_q=0\;;\; \forall \{\mu,\nu\} \subset \{1,2,3\} \;,\; \mathcal{T}_q=0$.\\

If we write all matrices $\rho'_j \; ,\; \rho''_j \;,\; \rho'''_j$ ; $\varrho_i \;,\; \varrho'_i\;,\; \tau_i\;,\;\tau'_i\;,\; \pi_i\;,\;\pi'_i$ ; $\rho_r \;,\; \rho_s \; ,\; \rho_q$ (as we have done for $\rho_j$ in Eq.(24)) in the Bloch representation and substitute them in Eq.(23) we get $\rho$ as in Eq.(19b).

To understand this let us see how the first term in Eq.(23) adds up to give the last term in Eq.(19b). The definition of $\rho_j,\;\rho'_j,\;\rho''_j,\;\rho'''_j$ (denoting the Bloch vectors by $s_1, s_2, s_3, s_4,....$) can be summarized in the tabular form\\

\vspace{.2in}

\bc
{\bf Table 1} \\

Correspondence between the first term in Eq.(23) and the last term in Eq. (19b).
\vspace{.2in}\\

\begin{tabular}{||c|c|c|c||c|c|c|c||}
\hline
&$s_1$ & $s_2$ & $s_3$ &$s_1s_2$&$s_1s_3$&$s_2s_3$&$s_1s_2s_3$\\
\hline
$\rho_j$& $\tilde{u}_j$&$\tilde{v}_j$ &$\tilde{w}_j$ & $\tilde{u}_j\tilde{v}_j$&$\tilde{u}_j\tilde{w}_j$&$\tilde{v}_j\tilde{w}_j$&$\tilde{u}_j\tilde{v}_j\tilde{w}_j$\\
\hline
$\rho'_j$& $\tilde{u}_j$&$-\tilde{v}_j$ &$-\tilde{w}_j$ & $-\tilde{u}_j\tilde{v}_j$&$-\tilde{u}_j\tilde{w}_j$&$\tilde{v}_j\tilde{w}_j$&$\tilde{u}_j\tilde{v}_j\tilde{w}_j$\\
\hline
$\rho''_j$& $-\tilde{u}_j$&$\tilde{v}_j$ &$-\tilde{w}_j$ & $-\tilde{u}_j\tilde{v}_j$&$\tilde{u}_j\tilde{w}_j$&$-\tilde{v}_j\tilde{w}_j$&$\tilde{u}_j\tilde{v}_j\tilde{w}_j$\\
\hline
$\rho'''_j$& $-\tilde{u}_j$&$-\tilde{v}_j$ &$\tilde{w}_j$ & $\tilde{u}_j\tilde{v}_j$&$-\tilde{u}_j\tilde{w}_j$&$-\tilde{v}_j\tilde{w}_j$&$\tilde{u}_j\tilde{v}_j\tilde{w}_j$\\
\hline
 \end{tabular}\\
 
\ec

\vspace{.2in}

The contribution of each column to  $\rho_j+\rho'_j+\rho''_j+\rho'''_j$ is zero except the last column which reproduces the last term in Eq.(19b). We can get the contributions of each term in $\rho_j,\;\rho'_j,\;\rho''_j,\;\rho'''_j$ to their sum by just keeping track of their signs. Thus we only need the following table (dropping $j$)

\vspace{.2in}
\bc
{\bf Table 2} \\

Contributions of various terms in $\rho, \rho', \rho'',\rho'''$  to their sum.
\vspace{.2in}\\

\begin{tabular}{||c|c|c|c||c|c|c|c||}
\hline
&$s_1$ & $s_2$ & $s_3$ &$s_1s_2$&$s_1s_3$&$s_2s_3$&$s_1s_2s_3$\\
\hline
$\rho$& $+$&$+$ &$+$ & $+$&$+$&$+$&$+$\\
\hline
$\rho'$& $+$&$-$ &$-$ & $-$&$-$&$+$&$+$\\
\hline
$\rho''$& $-$&$+$ &$-$ &$-$&$+$& $-$&$+$\\
\hline
$\rho'''$& $-$&$-$ &$+$&$+$&$-$&$-$&$+$\\
\hline
 
\end{tabular}\\

\ec
\vspace{.2in}

In the same way, the contributions of the terms involving $\varrho,\;\tau,\;\pi$ are obtained by using the table corresponding to table 2  for the bipartite case [9]. $\varrho,\;\tau,\;\pi$ which contain tensors of order two correspond to three 2-partite subsystems 12,13 and 23 . The corresponding tables are

\vspace{.2in}
\bc
{\bf Table 3} \\

Contributions to $\varrho+ \varrho'$
\vspace{.2in}\\
\begin{tabular}{||c|c|c|c||c|c|c|c||}
\hline
&$s_1$ & $s_2$ & $s_3$ &$s_1s_2$&$s_1s_3$&$s_2s_3$&$s_1s_2s_3$\\
\hline
$\varrho$& $+$&$+$ &$0$ & $+$&$0$&$0$&$0$\\
\hline
$\varrho'$& $-$&$-$ &$0$ & $+$&$0$&$0$&$0$\\
\hline
 
\end{tabular}\\

\ec

\vspace{.2in}

\bc
{\bf Table 4} \\

Contributions to $\tau+\tau'$
\vspace{.1in}\\

\begin{tabular}{||c|c|c|c||c|c|c|c||}
\hline
&$s_1$ & $s_2$ & $s_3$ &$s_1s_2$&$s_1s_3$&$s_2s_3$&$s_1s_2s_3$\\
\hline
$\tau$& $+$&$0$ &$+$ & $0$&$+$&$0$&$0$\\
\hline
$\tau'$& $-$&$0$ &$-$ & $0$&$+$&$0$&$0$\\
\hline
 
\end{tabular}\\

\ec

\vspace{.1in}

\bc
{\bf Table 5} \\
Contributions to $\pi+ \pi'$
\vspace{.2in}\\
\begin{tabular}{||c|c|c|c||c|c|c|c||}
\hline
&$s_1$ & $s_2$ & $s_3$ &$s_1s_2$&$s_1s_3$&$s_2s_3$&$s_1s_2s_3$\\
\hline
$\pi$& $0$&$+$ &$+$ & $0$&$0$&$+$&$0$\\
\hline
$\pi'$& $0$&$-$ &$-$ & $0$&$0$&$+$&$0$\\
\hline
 
\end{tabular}\\

\ec

\vspace{.2in}

Tables 2, 3, 4, 5 encode the procedure to construct the possible separable state given in Eq.(23).

 We now note the following points
\begin{verse}

(i) If the condition (21) holds, then the coefficient of the matrix $I_{d_1d_2d_3}$ in Eq.(23) is positive which ensures that the decomposition (23) of $\rho$ is positive semidefinite.\\

(ii) By virtue of Eq.(6), all the coherence vectors occurring in  $\rho'_j \; ,\; \rho''_j \;,\; \rho'''_j$ ; $\varrho_i \;,\; \varrho'_i\;,\; \tau_i\;,\;\tau'_i\;,\; \pi_i\;,\;\pi'_i$ ; $\rho_r \;,\; \rho_s \; ,\; \rho_q$ belong to the corresponding Bloch spaces.
\end{verse}

By (i) and (ii) we conclude that  $\rho'_j \; ,\; \rho''_j \;,\; \rho'''_j$ ; $\varrho_i \;,\; \varrho'_i\;,\; \tau_i\;,\;\tau'_i\;,\; \pi_i\;,\;\pi'_i$ ; $\rho_r \;,\; \rho_s \; ,\; \rho_q$ constitute density matrices. Further, all these matrices satisfy condition (11) so that, via proposition 1, all these matrices correspond to pure separable states, equal to the tensor products of their reductions. Therefore, they constitute density matrices and they are separable and so must be $\rho$. \hfill $\square$\\

We can generalize proposition 2 to the $N$-partite case by constructing the tables successively for $N=4,5,6,\cdots$. First note that the number of $\rho$ s in the first term of Eq.(23) lifted to the $N$-partite case is $2^{N-1}$. For $N=4$ we have eight. The corresponding table is

\vspace{.6in}
\bc
{\bf Table 6} \\

Generalization of Table 1 to $N=4$.
\vspace{.2in}\\

\begin{tabular}{||c|c|c|c|c||c|c|c|c|c|c|c|c|}
\hline
&$s_1$ & $s_2$ & $s_3$&$s_4$ &$s_1s_2$&$s_1s_3$&$s_1s_4$&$s_2s_3$&$s_2s_4$&$s_3s_4$&$s_1s_2s_3$&$s_1s_2s_4$\\
\hline
$\rho^{(1)}$& $+$&$+$ &$+$ & $+$&$+$&$+$&$+$& $+$&$+$ &$+$ & $+$&$+$\\
\hline
$\rho^{(2)}$& $+$&$+$ &$-$ & $-$&$+$&$-$&$-$& $-$&$-$ &$+$ & $-$&$-$\\
\hline
$\rho^{(3)}$& $+$&$-$ &$+$ & $-$&$-$&$+$&$-$& $-$&$+$ &$-$ & $-$&$+$\\
\hline
$\rho^{(4)}$& $+$&$-$ &$-$ & $+$&$-$&$-$&$+$& $+$&$-$ &$-$ & $+$&$-$\\
\hline
$\rho^{(5)}$& $-$&$+$ &$+$ & $-$&$-$&$-$&$+$& $+$&$-$ &$-$ & $-$&$+$\\
\hline
$\rho^{(6)}$& $-$&$+$ &$-$ & $+$&$-$&$+$&$-$& $-$&$+$ &$-$ & $+$&$-$\\
\hline
$\rho^{(7)}$& $-$&$-$ &$+$ & $+$&$+$&$-$&$-$& $-$&$-$ &$+$ & $+$&$+$\\
\hline
$\rho^{(8)}$& $-$&$-$ &$-$ & $-$&$+$&$+$&$+$& $+$&$+$ &$+$ & $-$&$-$\\
\hline

\end{tabular}\\

\ec
\vspace{.2in}
\bc
{\it (Table 6. Continued)}\\

\vspace{.1in}
\begin{tabular}{|c|c|c||}
\hline
$s_1s_3s_4$&$s_2s_3s_4$&$s_1s_2s_3s_4$\\
\hline
$+$&$+$& $+$\\
\hline
$+$&$+$& $+$\\
\hline
$-$&$+$& $+$\\
\hline
$-$&$+$& $+$\\
\hline
$+$&$-$& $+$\\
\hline
$+$&$-$& $+$\\
\hline
$-$&$-$& $+$\\
\hline
$-$&$-$& $+$\\
\hline

\end{tabular}\\

\ec
\vspace{.2in}

We see that the contribution of each column to the sum $\sum_i \rho^{(i)}$ is zero except the last one corresponding to the Kruskal decomposition of $\mathcal{T}^{(N)}$ occurring in the Bloch representation of the given state $\rho$. For general case of $N$-partite state we construct the table for $\rho^{(i)},\; i=1,2,\cdots,2^{N-1}$ as follows. First column consists of $2^{N-2}$ plus signs followed by  $2^{N-2}$ minus signs. Second column comprises alternating $2^{N-3}$ plus and minus signs. Continuing in this way upto $2^{N-N}=1$ we get alternating plus and minus signs in the $(N-1)$th column. We set the $N$th column to ensure that there are zero or even number of minus signs in each row. Rest  of the columns can be constructed by appropriate multiplications. This procedure can be checked on table 6. We denote the  sequence of such tables for $N=2,3,4,\cdots$ as $T_i,\; i=2,3,4,\cdots$.

The tables corresponding to $(N-1), (N-2),...,2$ partite subsystems giving rise to the remaining terms in the equation (23), lifted to $N$-partite case, are obtained from $T_{N-1},T_{N-2},...,T_3,T_2,$ exactly as described in the proof of  proposition 2. In this way we can lift eq.(23) to the $N$-partite case, with the total numbers of terms $\sum_{i=0}^{N-1}\binom{N}{i}2^{N-1-i} +1$. Once this is done, the rest of the proof for $N$-partite case follows as in proposition 2. Thus we have\\

\textbf{Proposition (2a)}: If a $N$-partite state $\rho$ acting on  $\mathcal{H}=\mathcal{H}^{d_1}\otimes \mathcal{H}^{d_2} \otimes \cdots \otimes \mathcal{H}^{d_N},\; d_1 \le d_2 \le \cdots \le d_N$ with  Bloch representation (8), where all $\mathcal{T}^{(k)},\; k>2$ have the completely orthogonal Kruskal decomposition, satisfy 

$$\sum_k\sq{\fr{2(d_k-1)}{d_k}} ||\mathbf{s_k}||_2+\sum_{\{\mu,\nu\}}\sq{\fr{4(d_{\mu}-1)(d_{\nu}-1)}{d_{\mu}d_{\nu}}} ||T^{\{\mu,\nu\}}||_{KF}$$

$$+\sum_{\{\mu,\nu,\kappa\}}\sq{\fr{8(d_{\mu}-1)(d_{\nu}-1)(d_{\kappa}-1)}{d_{\mu}d_{\nu}d_{\kappa}}} ||\mathcal{T}^{\{\mu,\nu,\kappa\}}||_{KF}+\cdots+$$
$$\sum_{\{k_1,k_2,\cdots,k_M\}}\sq{\fr{2^M\Pi_{k_i}(d_{k_i}-1)}{\Pi_{k_i}d_{k_i}}}||\mathcal{T}^{\{k_1,k_2,\cdots,k_M\}}||_{KF} +\cdots+\sq{\fr{2^N\Pi_{i}^N(d_{i}-1)}{\Pi_{i}^Nd_{i}}}||\mathcal{T}^{(N)}||_{KF} \le 1,  \eqno{(21)}$$\\

 then $\rho$ is separable.\hfill $\square$\\

For a $N$-qubit system  Theorem 1 and proposition 2a  together imply \\

 \textbf{Corollary 2 :} Let a $N$-qubit state have a Bloch representation

 $$\rho= \fr{1}{2^N}(\otimes_{k=1}^N  I_2^{(k)}+\sum_{\alpha_1 \cdots \alpha_N} t_{\alpha_1 \cdots \alpha_N} \la_{\alpha_1}^{(1)}\la_{\alpha_2}^{(2)} \cdots \la_{\alpha_N}^{(N)}),$$
 and let the tensor in the second term have the completely orthogonal Kruskal decomposition. 
 Then $\rho$ is separable if and only if $||\mathcal{T}^{(N)}||_{KF} \le 1.$\hfill $\square$\\

 
 
 \section{ Examples}

 We now investigate our separability criterion (18) for mixed states.     
  We consider $N$-qubit state $$\rho^{(N)}_{noisy}\;=\;\fr{1-p}{2^N}I+p|\psi\ran\lan\psi|, \;\;\;0\le p\le 1  \eqno{(25)}$$

where $|\psi\ran$ is a $N$-qubit $W$ state or GHZ state. We test for $N=3,4,5$ and $6$ qubits. We get,

\vspace{.2in}
\bc
{\bf Table 7} \\

The values of $p$ above which the states are entangled.
\vspace{.1in}\\

\begin{tabular}{||c|c||c||}
\hline
$|GHZ\ran$ & $|W\ran$ & $N$\\
$p >$  & $p >$  & \\ 
\hline
 0.35355 & 0.3068 & 3\\
\hline
 0.2 &  0.3018 & 4\\
\hline
 0.17675 &  0.30225 & 5\\
\hline
 0.1112 &  0.3045  & 6\\
\hline
\end{tabular}\\
\vspace{.2in} 

 \ec 
\vspace{.2in}   

  Entanglement in various partitions of $W$ noisy state Eq.(25) is obtained by using $(N-n)$ qubit reduced $W$ noisy state 
  $$\rho_{noisy}^{(N-n)}(W)=\fr{1-p}{2^{N-n}} I_{N-n} +\fr{n}{N} p |0_{N-n}\ran \lan 0_{N-n}|+\fr{N-n}{N} p |W_{N-n}\ran \lan W_{N-n}|  \eqno{(26)}$$
  
    For $N=6$ and $n=2$ we found that the state is entangled for $0.491 < p \le 1.$   

  For $N$ qutrits $(d=3)$ we test for $$\rho_{noisy}^{(N)}\;=\;\fr{1-p}{3^N}\;I\;+\; p|\psi\ran\lan\psi|\eqno{(27)}$$ where    $|\psi \ran = \fr{1}{\sqrt{d}}\sum_{k=1}^d |kkk \dots  \ran $ is the maximally entangled state for $N$ qutrits.\\
   
   For $N=3$ and $N=4$ (qutrits) the state $\rho_{noisy}^{(N)}$ in Eq. (27) is entangled for 
   
   $$0.2285 < p \le 1,  \; \; \; \; N=3$$
   
    $$0.2162 < p \le 1, \; \; \; \; N=4  \eqno{(28)}$$

  The state $$\rho_{noisy}\;=\;\fr{1-p}{24}\;I\;+\; p|\psi\ran\lan\psi| \eqno{(29)}$$  where  $|\psi\ran=\fr{1}{2}(|112\ran+|123\ran+|214\ran+|234\ran) $ in the space $\mathbb{C}^2\otimes\mathbb{C}^3\otimes \mathbb{C}^4$ is found to be entangled for $0.24152 < p\le 1.$\\
        
   All of the above examples involve NPT states. Now we apply our criterion to PPT entangled states for which PPT criterion is not available.\\
  
  We apply our criterion to the three qutrit bound entangled state considered by L. Clarisse and P. Wocjan [27], given by $\rho_c \otimes|\psi\ran \lan \psi|$ where $\rho_c$ is the chess-board state given in [27] and $|\psi\ran$ is an uncorrelated ancilla. Our criterion detects the entanglement of this state as $||\mathcal{T}^{(12)}|| =3.75 >3.$
  Further, the four qutrit state $\rho = (1-\beta)\rho_c \otimes \rho_c + \beta I/81$ considered by the same authors yields entanglement for $0 \le \beta \le 0.2,$ after tracing out either subsystems 1 and 2 or subsystems 3 and 4.
  
  Now we consider the important example of the Smolin state [28,29], which is a four qubit bound entangled state given by $$ \rho_{ABCD}^{unlock}=\frac{1}{4}\sum_{i=1}^4 |\psi_{AB}^i\ran \lan \psi_{AB}^i|\otimes |\psi_{CD}^i\ran \lan \psi_{CD}^i| \eqno{(30)}$$   
  
  where $|\psi_{AB}^i\ran$ and $|\psi_{CD}^i\ran$ are the Bell states. $ \rho_{ABCD}^{unlock}$ has the Bloch representation $\rho_{ABCD}^{unlock}=\frac{1}{16}(I^{\otimes4}+\sum_{i=1}^3 \si_i^{\otimes4})$ so that Corollary 2 applies (note that the requirement of completely orthogonal Kruskal decomposition is trivially satisfied). We find for this state $||\mathcal{T}^{(4)}||_{KF} = 3 > 1$ confirming its entanglement.\\
    
  Our last example is the four qubit bound entangled state due to W. D\"ur [30,31]
  $$\rho_4^{BE}= \frac{1}{5}(|\psi\ran \lan \psi|+\frac{1}{2}\sum_{i=1}^4(P_i+\overline P_i))$$   
  where $|\psi\ran$ is a 4-party (GHZ) state , $P_i$ is the projector onto the state $|\phi_i\ran$, which is a product state equal to $|1\ran$ for party $i$ and $|0\ran$ for the rest , and $\overline P_i$ is obtained from  $P_i$ by replacing all zeros by ones and vice versa. We get $||\mathcal{T}^{(4)}||_{KF} = 1.4 > 1$ confirming the entanglement of this state. \\

  
\section{Summary}

  In conclusion we have presented a new criterion for separability of $N$ partite quantum states based on the Bloch representation of states. This criterion is quite general, as it  applies to all $N$-partite quantum states living in $\mathcal{H}=\mathcal{H}^{d_1}\otimes \mathcal{H}^{d_2} \otimes \cdots \otimes \mathcal{H}^{d_N},$ where, in general, the Hilbert space dimensions of various parts are not equal. Most of the previous such criteria had restricted domain of applicability like the states supported on symmetric subspace [4] or, are, in general, restricted to bipartite case.
   In proposition 2 we have given a sufficient condition for the separability of a tripartite state under the condition that the tensors occurring in the Bloch representation of the state have completely orthogonal Kruskal decomposition. This result can be generalized to the $N$-partite case. Via corollary 2 we give a necessary an sufficient condition to test the separability of a class of $N$-qubit states which includes $N$-qubit PPT states. Smolin state (30) is an important example in this class. The key idea in our work is the matrization of multidimensional tensors, in particular, Kruskal matrization. We have defined a new tensor norm as the maximum of the KF norms of all the matrix unfoldings of a tensor, which is easily computed. We have also shown that this norm can be calculated even more efficiently  for a $N$-qudit state supported in the symmetric subspace. 
   It will be interesting to seek a relation of this tensor norm with other entanglement measures. Again, the entanglement measures like concurrence known so far are successfully applied to pure states, bipartite or multipartite, while our tensor norm can be easily computed for arbitrary $N$-partite quantum state. Finally, our result on full separability (proposition 1) of $N$-partite pure states can be easily moulded for the $k$-separability of an $N$-partite pure state. In fact it is straightforward to construct an algorithm giving the complete factorization of the $N$-partite pure state (see the paragraph following the proof of proposition 1). It is also easy to see that theorem 1 can be applied to any partition of a $N$-partite system via the Bloch representation in terms of the generators of the appropriate $SU$ groups. Most important is the observation that all the tensors in the Bloch representation can be computed using the measured values of the basis operators $\{\lambda_{\alpha_k}\}$ so that our detectiblity criterion is experimentally implementable.\\

\vspace{0.3cm}

{\bf Acknowledgments} \\

 We thank Guruprasad Kar and R.Simon for encouragement. We thank Julio I. de Vicente for a very useful communication.  We thank Guruprasad Kar and Sibasish Ghosh for suggesting the last two examples. ASMH thanks Sana'a University for financial support. \\

{\bf References}

\begin{verse}

[1] M. B. Plenio and S. Virmani (2007), {\it An introduction to entanglement measures}, Quantum Inf. Comput., Vol.  \textbf{7}, pp. 001-051.

[2] K. \.Zyczkowski and I. Bengstsson (2006),  {\it An introduction to quantum entanglement: a geometric approach},  quant-ph/0606228.

[3] R.F. Werner (1989), {\it Quantum states with Einstein-Podolsky-Rosen correlations admitting a hidden-variable model }, Phys. Rev. A , \textbf{40}, pp. 4277-4281.

[4] A. R. Usha Devi, R. Prabhu, and A. K. Rajagopal (2007), {\it Characterizing Multiparticle Entanglement in Symmetric N-Qubit States via Negativity of Covariance Matrices
}, Phys. Rev. Lett., \textbf{98}, pp. 060501.

[5] Florian Mintert, Marek Ku\'s, and Andreas Buchleitner (2005), {\it Concurrence of mixed multipartite quantum states}, Phys. Rev. Lett., \textbf{95}, pp. 260502. 

[6] K. Chen and L. Wu (2002), {\it The generalized partial transposition criterion for separability of multipartite quantum states}, Phys. Lett. A, \textbf{306}, pp. 14-20. 

[7] A. Ac\'in, D. Bruss, M. Lewenstein and A. Sanpera  (2001), {\it Classification of mixed three-qubit states 
}, Phys. Rev. Lett., \textbf{87}, pp. 040401 .

[8] W. D\"ur, J. I. Cirac, and R. Tarrach (1999), {\it Separability and distillability of multiparticle quantum systems }, Phys. Rev. Lett., \textbf{83}, pp. 3562. 

[9]  Julio I. de Vicente (2007), {\it Separability criteria based on the Bloch representation of density matrices }, Quantum Inf. Comput., Vol. \textbf{7}, pp. 624-638.

[10] L. De Lathauwer, B. De Moor, and J. Vandewalle (2000), {\it A multilinear singular value decomposition
}, SIAM J. Matrix Anal. A., Vol. \textbf{21}, pp. 1253-1278.

[11] L. De Lathauwer, B. De Moor, and J. Vandewalle (2000), {\it On the best rank-1 and rank-$(R_1,R_2, . . . , R_N)$
approximation of higher-order tensors}, SIAM J. Matrix Anal. A., Vol. \textbf{21}, pp. 1324-1342.

[12] T. G. Kolda (2006), {\it Multilinear operators for higher-order decompositions}, Tech. Report SAND2006-2081, Sandia National Laboratories, Albuquerque, New Mexico and
Livermore, California, Apr. 2006.

[13] T. G. Kolda (2001), {\it Orthogonal tensor decompositions}, SIAM J. Matrix Anal. A., Vol. \textbf{23}, pp. 243-255.

[14] J. B. Kruskal  (1977), {\it Three-way arrays: rank and uniqueness of trilinear decompositions,
with application to arithmetic complexity and statistics}, Linear Algebra Appl., Vol. \textbf{18}, pp. 95-138.

[15] T. Zhang and G. H. Golub (2001), {\it Rank-one approximation to high order tensors}, SIAM J. Matrix Anal. A., Vol. \textbf{23}, pp. 534-550.

[16] B. W. Bader and T. G. Kolda (2004), {\it Matlab tensor classes for fast algorithm prototyping}, Tech. Report SAND2004-5187, Sandia National Laboratories, Albuquerque,
New Mexico and Livermore, California, Oct. 2004. 

[17]  Brett W. Bader and Tamara G. Kolda (2006), {\it Efficient matlab computations with sparse and factored tensors}, REPORT SAND2006-7592 Sandia National Laboratories, Albuquerque, New Mexico 87185 and Livermore, California 94550, Printed December 2006.

[18] F. Bloch (1946), {\it Nuclear induction}, Phys. Rev., \textbf{70}, pp. 460-474.

[19] G. Kimura and A. Kossakowski  (2005), {\it  The Bloch-vector space for N-level systems-the spherical-coordinate point of view }, Open Sys. Inf. Dyn., Vol. \textbf{12}, pp. 207.

[20] G. Kimura (2003), {\it The Bloch vector for N-level systems}, Phys. Lett. A, \textbf{314}, pp. 339-349.

[21] M.S. Byrd and N. Khaneja (2003), {\it Characterization of the positivity of the density matrix in terms of the coherence vector representation }, Phys. Rev. A, \textbf{68}, pp. 062322-062335.

[22] G. Mahler and V.A. Weberru\ss (1995), {\it Quantum Networks}, Springer (Berlin).

[23] J. E. Harriman (1978), {\it Geometry of density matrices. I. Definitions, N matrices and 1 matrices }, Phys. Rev. A, \textbf{17}, pp. 1249-1256.

[24] A. Kossakowski (2003), {\it A class of linear positive maps in matrix algebras}, Open Sys. Inf. Dyn., Vol. \textbf{10}, pp. 1.

[25] This example is taken from ref.[10],  we recommend [10] and [16] to clarify basic concepts.

[26] R.A. Horn and C.R. Johnson (1991), {\it Topics in matrix analysis}, Cambridge University Press (Cambridge).

[27] L. Clarisse and P. Wocjan (2006), {\it On independent permutation separability criteria}, Quantum Inf. Comput., Vol. \textbf{6}, pp. 277-288.\\

[28] R., P., M., K. Horodecki, {\it  Quantum Entanglement }, quant-ph/0702225v2.\\

[29] John A. Smolin (2001), {\it Four-Party Unlockable Bound Entangled State }, Phys. Rev. A, \textbf{63}, pp. 032306-032310.\\

[30] W. D\"ur (2001), {\it Multipartite Bound Entangled States that Violate Bell's Inequality }, Phys. Rev. Lett., \textbf{87}, pp. 230402.\\

[31] A. Ac\'in (2002), {\it Distillability, Bell Inequalities, and Multiparticle Bound Entanglement }, Phys. Rev. Lett., \textbf{88}, pp. 027901.\\

\end{verse} 

\end{document}